\documentclass[nohyperref]{article}

\usepackage{microtype}
\usepackage{graphicx}
\usepackage{subfigure}
\usepackage{booktabs} 

\usepackage{hyperref}

\usepackage[preprint]{icml2023}

\usepackage{amsmath}
\usepackage{amssymb}
\usepackage{mathtools}
\usepackage{amsthm}

\usepackage{algorithmic}
\usepackage{textcomp}
\usepackage{xcolor}

\usepackage{fancyhdr}
\usepackage[normalem]{ulem}
\usepackage{pifont}

\usepackage{comment} 
\usepackage{tabularx,tabulary} 
\usepackage{multirow}
\usepackage{xspace} 
\usepackage{enumitem}
\usepackage{soul} 
\usepackage{xspace}

\newcommand{\josep}[1]{{\footnotesize\color{orange}[Josep: #1]}}

\usepackage[capitalize,noabbrev]{cleveref}

\theoremstyle{plain}

\theoremstyle{definition}

\theoremstyle{remark}

\usepackage[textsize=tiny]{todonotes}

\newcommand{\modelname}[1]{{{{\textit{DefenderGAN}}\xspace}}{#1}}

\icmltitlerunning{Defensive ML: Defending Architectural Side-channels with Adversarial Obfuscation}

\begin{document}

\twocolumn[
\icmltitle{Defensive ML: Defending Architectural Side-channels with Adversarial Obfuscation}




\begin{icmlauthorlist}
\icmlauthor{Hyoungwook Nam}{uiuc}
\icmlauthor{Raghavendra Pradyumna Pothukuchi}{yale}
\icmlauthor{Bo Li}{uiuc}
\icmlauthor{Nam Sung Kim}{uiuc}
\icmlauthor{Josep Torrellas}{uiuc}
\end{icmlauthorlist}

\icmlaffiliation{uiuc}{University of Illinois at Urbana-Champaign, USA }
\icmlaffiliation{yale}{Yale University, USA }

\icmlcorrespondingauthor{Hyoungwook Nam}{hn5@illinois.edu}
\icmlcorrespondingauthor{Josep Torrellas}{torrella@illinois.edu}

\icmlkeywords{Machine Learning, ICML, Adversarial Machine Learning, Deep Learning, GAN, Security, Computer Architecture}

\vskip 0.3in
]



\printAffiliationsAndNotice{\icmlEqualContribution} 

\begin{abstract}
Side-channel attacks that use machine learning (ML) for signal analysis have become prominent threats to computer security, as ML models easily find patterns in signals. 
To address this problem, this paper explores using Adversarial Machine Learning (AML) methods as a defense at the computer architecture layer to obfuscate side channels. 
We call this approach {\em Defensive} ML, and the generator to obfuscate signals, {\em defender}.
Defensive ML is a workflow to design, train, and deploy defenders for different environments.
First, we design a defender architecture given the physical characteristics and hardware constraints of the side-channel. 
Next, we use our DefenderGAN structure to train the defender.
Finally, we apply defensive ML to thwart two side-channel attacks: one based on memory contention  and the other on application power.
The former uses a hardware defender with ns-level response time that attains a high level of security with half the performance impact of a traditional scheme; the latter uses a software defender with ms-level response time that provides better security than a traditional scheme with only 70\% of its power overhead. 

\end{abstract}

\section{Introduction}
\label{intro}
Securing computers against information leakage has never been more challenging. 
Attackers are using complex analyses to 
exfiltrate sensitive information from a variety of \emph{side-channels}.
Researchers are finding out that microarchitectural components are potential side-channels for leaking secrets.
Shared architectural resources can be exploited by attackers to indirectly observe activities of victim applications that are supposed to be invisible to others.
An attacker can decode the observed activity patterns to infer the secret information.
Using machine learning (ML) can amplify the information leakage from the observed patterns.
Such ML based attacks can also resist basic defenses like noise addition so that thwarting ML attacks often require bigger distortions that result in high performance overheads.

Therefore, it is necessary to develop a more intelligent defense strategy that is effective at neutralizing many types of ML side-channel attacks.
An intriguing approach is to use adversarial machine learning (AML) methods for defense.
AML methods have been successful at creating adversarial examples~\cite{adversarialexample, advgan} that confuse ML classifiers to incorrectly recognize the inputs while minimizing the amount of noise to be added.
We propose using an ML noise generator to efficiently obfuscate an architectural side-channel, namely \emph{defender}.

To this end, we define the side-channel obfuscation as an ML problem.
Given the observed side-channel signals as inputs and corresponding secret labels as targets, a defender's objective is to 1) randomize the classification results of attackers and 2) minimize the noise level.
While the idea is straightforward, there are multiple challenges to this approach.
First, we need to minimize the hardware installation cost of the defender.
The budget for installing the ML defender can be very limited as it may need to operate at nanosecond-level.
Next, the defender must comply with physical limitations of the noise injection methods.
Most of all, a defender must confuse every possible classifier 
that the attacker can leverage.
That is, the defender's noise must be \emph{transferable} to a variety of classifiers.

We propose \emph{Defensive ML}, a workflow of designing, training, evaluating, and deploying defenders for architectural side-channels.
Given a side-channel, the first step is to design a defender network complying with the hardware cost and physical constraints of the side-channel.
Then we train the defender using \modelname{}, a DNN structure that adversarially trains a defender against a classifier and a discriminator.
Inspired by Adversarial GAN~\cite{advgan}, \modelname{} is designed to maximize the transferability of the defender to unknown black-box classifiers.
The trained defender is evaluated by its transferability to adaptive attackers that are trained in the presence of the defender's noise.
The defender with the best transferability is chosen to be deployed and compressed by GAN compression~\cite{gancompression} if needed.

Finally, we demonstrate Defensive ML by applying it to thwart two important and different side-channel attacks: one
exploiting memory contention and the other analyzing a computer's power consumption. 
The former 
uses a hardware defender with ns-level response time that
attains a high level of security with half the performance impact of a traditional scheme;
the latter uses a software defender 
with ms-level response time that provides better security than a traditional scheme with only 70\% of its power 
overhead.
The contributions of this paper are: 
\begin{itemize}
\item Identifying side-channel obfuscation as an optimization problem and propose using ML defenders to mitigate architectural side-channels.
\item Proposing a workflow to design, train, evaluate, and deploy defenders against architectural side-channels.
\item Applying Defensive ML to thwart two side-channel attacks based on memory contention and 
application power, and demonstrating its efficacy.
\end{itemize}

\section{Background}
\label{back}
\subsection{Microarchitectural Side-channel Attack}

\begin{figure}[ht]
\centering
\includegraphics[width=\linewidth]{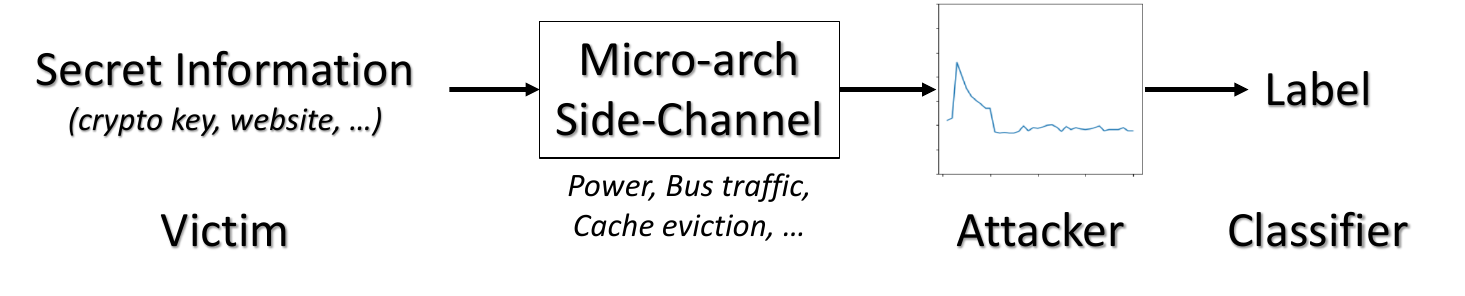}
\caption{A typical model of architectural side-channel attacks.}
\label{fig:sidechannel}
\end{figure}

A typical architectural side-channel attack with ML is illustrated at Figure~\ref{fig:sidechannel}.
First, a secret information can be leaked if the victim application's activity depends on the information.
Next, an attacker can indirectly observe such activity via shared resources like caches~\cite{cacheTut}, branch predictor~\cite{branchAttack}, PCI express~\cite{tan2021invisible}, and on-chip interconnects~\cite{lordofthering}.
Finally, the attacker can decode the observed activity patterns to infer the secret information like cryptographic keys~\cite{emAttack} and accessed websites~\cite{websiteAttackModel}.

Recently, attackers have begun using ML for side-channel analysis~\cite{batteryAttack, lordofthering}.
ML attacks not only amplify leakage from unprotected channels~\cite{crossDL, maghrebi2016breaking, lerman2015template}, but also circumvent defenses like noise addition, masking, and shuffling~\cite{perin2018lowering, cnnjitter}.


\subsection{Adversarial Attacks}
\label{aml}

Recent studies have found that ML models are vulnerable to attacks.
It is possible to generate \textit{adversarial examples}~\cite{adversarialexample} against 
an ML classifier which,
with very small perturbations to the inputs (even imperceptible to humans), drastically change the classification outcomes.

Sometimes, an adversarial attacker may target a black-box classifier whose parameters are not known at the time of the attack.
Fortunately, adversarial examples are often transferable~\cite{transferability} in that an attack targeting one classifier can be effective at another classifier.
One possible way to automatically generate such adversarial perturbations is training an ML generator that generates the adversarial examples.
Adversarial GAN (AdvGAN)~\cite{advgan} is an example of this method to train the generator.
AdvGAN adversarially trains the generator using the generative adversarial network (GAN)~\cite{gan} structure, amplifying its efficacy for black-box attacks.

\subsection{Adversarial Attack for Side-channel Defense}

There are two prior works that use adversarial attacks
as a countermeasure against side-channel attacks.
However, they cannot be considered as complete practical solutions due to their limitations.
Picek et al.~\cite{advattackposter} considered the power signals 
created by various plaintexts and analyzed them with  
ML attackers. Then, they showed that, with small, adversarial modifications to the trace, they can 
mislead the attackers. However, they only use classifiers trained with unprotected signals, which is
unrealistic as attackers can always train classifiers with the presence of the protection.

Rahman et  al.~\cite{mockingbird} showed the use of AML method against website fingerprinting 
attacks through traffic measurement.
Their defense is not very effective, in that  attackers can still achieve 38--42\% accuracy,
while a perfect defense of website fingerprinting with 95 classes 
 allows attackers to reach only 1.05\% accuracy.

Most of all, these systems require the full trace of a signal to produce the perturbations for the signal. As a result, they are post-processing solutions, not implementable to defend systems in the real world; they are proofs of concept.
A practical solution must be a dynamic defense, which produces the perturbations on the fly.

\section{Side-channel Defense Problem}
\label{lands}

Defending a side-channel is not just "attacking the attacker".
An adversarial attack is successful if it can confuse one target classifier of interest.
However, this is not a successful defense because the attacker can always leverage another classifier that works better.
Instead, an ML defender should minimize the \emph{information leakage} through the channel along with the overhead caused by the generated noise.

Assume that the attacker can observe a set of signals $X$ from the side-channel, where each signal $x\in X$ has corresponding secret information $s(x)$.
The ML defender $g$ obfuscates the channel, yielding the set of perturbed signals $X' = \{x + g(x) | x \in X\}$.
An attacker tries to reconstruct the secret information using an ML classifier $f$, yielding the classification results $f(X') = \{f(x+g(x)) | x \in X\}$.
The information leakage (channel capacity) can by quantified by the mutual information between the secret information and the classification results, $I(f(X');S)$, where  $S = \{s(x)|x\in X\}$ is the set of secret information $s$.
The leakage is zero if the distribution of $f(X')$ is independent to the distribution of $S$.
That is, a well-protected side-channel makes classification results no better than random guesses.
Hence, one way to optimize the defender for minimal information leakage is setting the classification target of perturbed signals to a uniform distribution.

The defender must resist \emph{all} possible classifiers in that the capacity of the obfuscated side-channel must be determined by the best possible classifier.
However, it is impossible to find the absolute best as there are infinitely many possible classifiers.
Instead, we compute the empirical expected channel capacity $max_{i} I(f_i(X');S)$ by using a finite set of classifiers $f_1 \dots f_N$.
While this empirical capacity does not guarantee the absolute security level of the defended side-channel, it provides a good estimation of transferability of the defender to various types of ML classifiers.

\begin{figure}[t]
\centering
\includegraphics[width=0.9\linewidth]{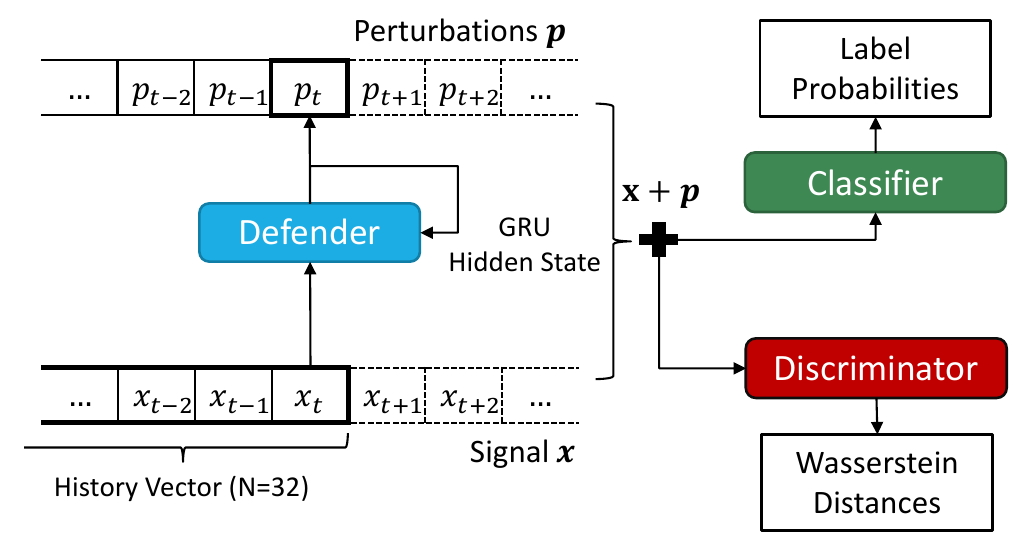}
\caption{Training of the  networks in the \modelname{}.}
\label{fig_architecture}
\vskip -0.1in
\end{figure}

There are multiple challenges to be addressed for implementing an ML defender to microarchitectural side-channels.
First, we need to consider the hardware implementation cost.
Some side-channels such as those arising from the cache and interconnect can operate at the nanosecond timescale.
In those cases, the defender must be compressed aggressively to reduce its cost.
Next, the defender must comply with the physical limitations of noise injection methods.
For example, many side-channels are based on latency/timing signals measured in cycle counts.
A defender for a timing side-channel can only add positive integer latency for obfuscation since it cannot reverse the time to reduce the latency.
Most of all, an attacker can try one's best to choose the most effective classifier for decoding the side-channel.
That is, the defender's noise must be \emph{transferable} to every possible classifier, regardless of its method.
Especially, the defender must resist \emph{adaptive attacks} where the attackers try to train the classifier with the presence of the generated noise.

\section{Defensive ML}
\label{amldefense}

\subsection{Designing the Defender Network}
\label{sec:defender}

A defender is an ML model that determines the current noise to add given the measurements of previous signals.
The implementation method of the defender depends on the timing and frequency requirements of the target side-channel.
For example, it is possible to implement the defender by software if the side-channel operates at resolution of milliseconds.
On the other hand, some architectural side-channels can run at much faster rate, even once in a few nanoseconds.
In those cases, it is necessary to integrate the defender to the hardware design to meet the timing requirements.

The hardware installation cost is the main factor that determines the allowable size of the neural network.
Contemporary neural network accelerator designs can achieve up to 10 TOPS/W~\cite{reuther2019survey} so that one can estimate the maximum amount of computation per inference given the power budget and timing requirements.
For instance, assume that the power budget is 1W and the defender needs to run at 500MHz.
Then, the computation per defender inference must not exceed 20k operations = $(10 TOPS/W) \times (1W) / (0.5GHz)$.

Most side-channel signals are sequences so that it is desirable to use an ML model that catches sequential dependencies.
Recurrent neural networks (RNN) are good candidates because they can handle sequential dependencies with memory states and tend to have low inference cost.
Transformers and convolution neural networks can also manage sequential patterns but they are not suitable for most cases due to their high computational costs.

\subsection{Training the Defender in \modelname{}}
\label{sec:defendergan}


Once we have designed the defender and collected labeled signals from the side channel,
we train the defender using proposed \modelname{} network.
Figure~\ref{fig_architecture} shows how we train the networks in the \modelname{}.
It follows the design principles of AdvGAN, but designed to solve the side-channel defense problem, instead of adversarial attack.

The first difference is that the defender must generate the noise on-the-fly so that it can only leverage previous history of the signal when determining the current noise level.
Specifically, at step $t$, it is fed the most
recent 32 samples \{$x_t$, $x_{t-1}$..\} and produces the perturbation $p_t$.
The resulting noisy trace $x+p$ is passed to the classifier and discriminator as an attacker can collect the signal and decode it offline.

Second, the defender's objective is to fully randomize the classifier's predictions, not confusing it to a specific target.
Therefore, when training the defender, the classification target of the classifier is the uniform probability distribution instead of a single fake label.
In binary classification, we use the flipped target $(0\rightarrow 1, 1\rightarrow 0)$ instead, because we empirically find that the flipped target is better against binary classifiers.
The defender is evaluated by how close the predictions are to the random guesses.

Next, unlike in an AdvGAN, the classifier is also adversarially trained against the defender using perturbed signals generated by it.
The reason is that the defender must resist adaptive attacks that are trained in presence of the noisy signals.
Also, this adversarial training with the dynamically adapting classifier helps the defender being transferable to other adaptive attackers.

Finally, the discriminator does not distinguish real and fake signals as there is no such distinction after deploying the defender.
Instead, the discriminator acts like another classifier which gives a different type of adversarial pressure to the defender.
While the classifier minimizes the cross entropy between its predictions and the true labels, the discriminator tries to maximize Wasserstein distance~\cite{wgan} between label distributions.

The defender $g$ generates the perturbation $p=g(x)$, and  the classifier and the discriminator are trained with the perturbed signal $x+p$.
The classifier $f$ is trained with the cross-entropy loss $L_C(y,f(x+p))$ between its prediction and the true label $y$.
The discriminator $d$ is optimized to minimize our \emph{one-to-all} Wasserstein loss $L_D(y,d(x+p))$.
While the original Wasserstein loss can only handle binary cases (real or fake), the one-to-all can manage multi-label cases.
With the discriminator outputs $d(x+p) = \{d^{(1)}(x+p), \dots d^{(n)}(x+p)\}$, the one-to-all loss $L_D$ is defined as below.
$$L_D(y,d(x+p)) = \mathbb{E}_{i\neq y}(d^{(i)}(x+p)) - d^{(y)}(x+p)$$
Minimizing $L_D$ should maximize the one-to-all distances between label distributions.
Like in Wasserstein GAN, we apply weight clipping to the parameters with the threshold of $+-0.01$.
The defender's loss $L_G$ has three components: (1) the Kullback-Leibler divergence $kl(\cdot,\cdot)$ between the classifier output probabilities $f(x+p)$ and the uniform probability $u$ (inverted probability in binary case), (2) the negated discriminator loss $L_D$ to minimize the distances, and (3) the hinged L1-norm of the perturbation output.
$$ L_G = kl(f(x+p),u) - \lambda_d L_D + \lambda_h max(0, ||p|| - P)$$
By adjusting the value of $P$, we can control the perturbation level of the defender.
The detailed implementation and hypereparameters ($\lambda_d, \lambda_h$) are available at the source code in the supplementary materials.

\subsection{Assessing Defender Transferability}
\label{sec:adaptive}
\begin{figure}[ht]
\centering
\includegraphics[width=0.85\linewidth]{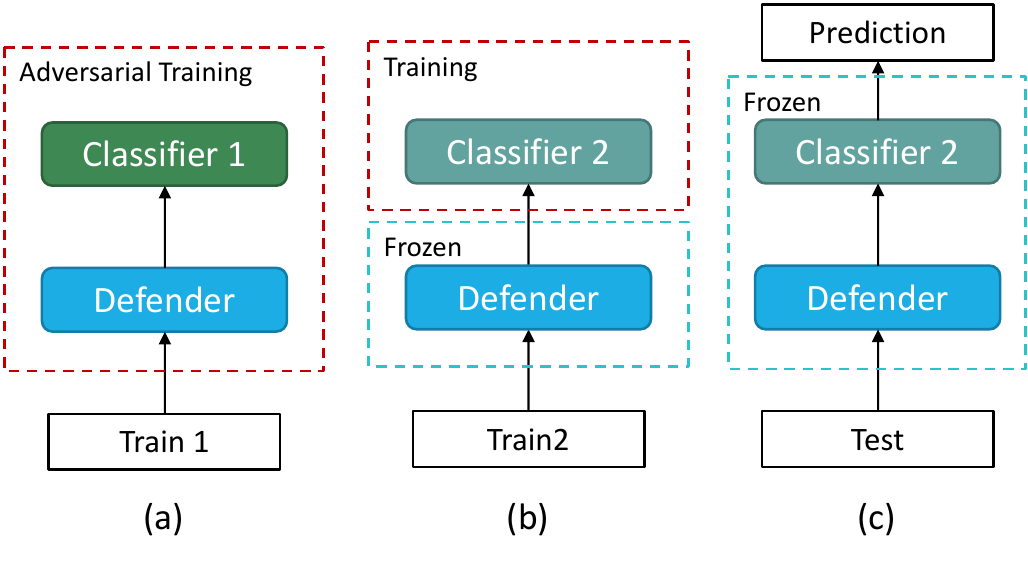}
\caption{Training and deploying a defender.}
\label{fig_traintest}
\vskip -0.1in
\end{figure}

Figure~\ref{fig_traintest} illustrates the procedure of testing the transferability of the defender against adaptive attacks.
The defender can only be adversarially trained with a certain classifier (Classifier 1) on a certain training data (Train 1).
Then, we emulate the adaptive attack using a new classifier (Classifier 2) and a new training data (Train 2).
Here, Classifier 2 is trained with the presence of the defender's noise and the defender's parameters are frozen as we cannot update the defender after deployment.
Finally, we test the efficacy of Classifier 2 at a new environment (Test).

We select various types of ML classifiers for Classifier 2 and, for each of them, perform the steps shown in Figures~\ref{fig_traintest}b and c.
The best performing classifier's test accuracy is reported, which is the worst case for the defender.
This determines the expected channel capacity of the defended side-channel.

\subsection{Deploying the Defender}
\label{sec:deploy}

After training and evaluating the defender, the defender with the best transferability is chosen and deployed for the side-channel obfuscation.
As discussed above, the defender may be deployed via software or hardware depending on the characteristics of the side-channels.
The defender may additionally have to be compressed for a hardware deployment.
This compression can be performed using the GAN compression technique~\cite{gancompression}.
Given the bigger \emph{teacher} model, the compressed \emph{student} can be trained using both the knowledge distillation loss~\cite{distillation} and \modelname{}'s loss functions simultaneously.
We can quantize the compressed model using smaller precision numbers to further reduce the hardware cost.

\section{Applying Defensive ML to Side-channels}
\label{sec:sidechannel}

An effective design for a defender depends on the 
particular side channel and  victim applications that are being targeted.
We examine two case studies: memory contention and application power side channels. 
In this section, we outline the two cases and
describe our proposed defensive ML workflow, emphasizing the differences
between them.

The memory contention side-channel is an example of fast-paced microarchitectural channels that require a compressed hardware module to deploy the defender.
We provide trace-level experiments to show that it is possible to obfuscate the channel with a tiny compressed ML defender.
The application power side channel is a slower-paced side channel where it is possible to apply the defender in software.
We conduct real-system experiments to prove that defensive ML can practically defend against the channel.

\subsection{Defending Memory Contention Side Channel}

\begin{figure}[ht]
\centering
\includegraphics[width=0.9\linewidth]{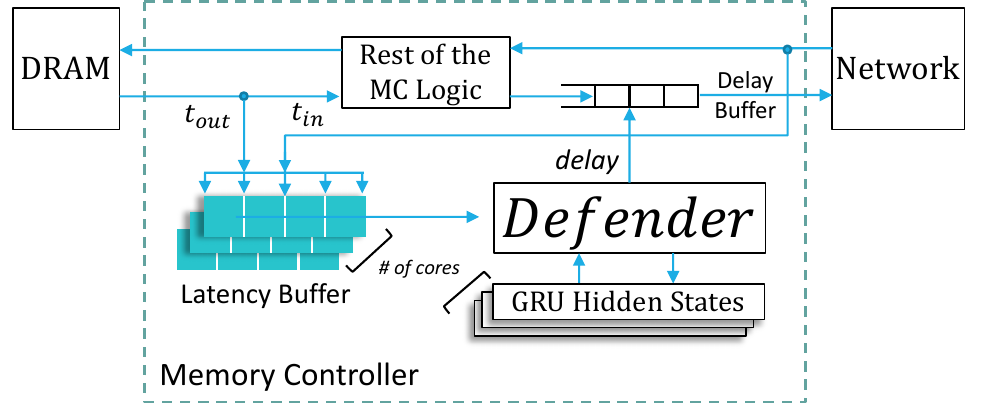}
\caption{Defender for the memory contention side channel.}
\label{fig_defense}
\end{figure}

The memory contention side-channel can leak information of a victim application through memory latency patterns.
We consider a chip multiprocessor where the victim application runs on one core and the attacker on another.
The victim's execution path depends on a secret bit; the attacker repeatedly accesses the main
memory and measures the latencies.
From the pattern of access latencies, an attacker can deduce the secret bit.
This attack is is based on the work of Paccagnella et al.~\cite{lordofthering}.


In this attack, we run iterative functions from two crypto applications: the RSA~\cite{rsa} and the
EDDSA~\cite{eddsa} decryption algorithms.
The victim applications run in loops, where each iteration processes one bit.
The samples of memory access latencies during an iteration form an input signal; the 
corresponding secret bit of the iteration  is the target label.

The envisioned defender is in a hardware unit in the memory controller (MC) that records the time
when a load arrives at the MC and the time when the main memory produces the requested data. To obfuscate the
latency of the load access, the defender stalls the returning data for a certain time period.
The defender cannot know which memory transaction is issued by the attacker so that it needs to process every load arriving to the MC.
In modern dual-channel DRAM (DDR4-3200), the fastest rate of memory transactions is once in every 1.25 nanoseconds.
Such a fast hardware module needs to consume reasonable power and area.
We believe that 1\% of a contemporary CPU's power and area are justifiable to invest for a side-channel defense, which are 1W and 1mm$^2$, respectively.

\begin{figure}[ht]
\centering
\includegraphics[width=0.9\linewidth]{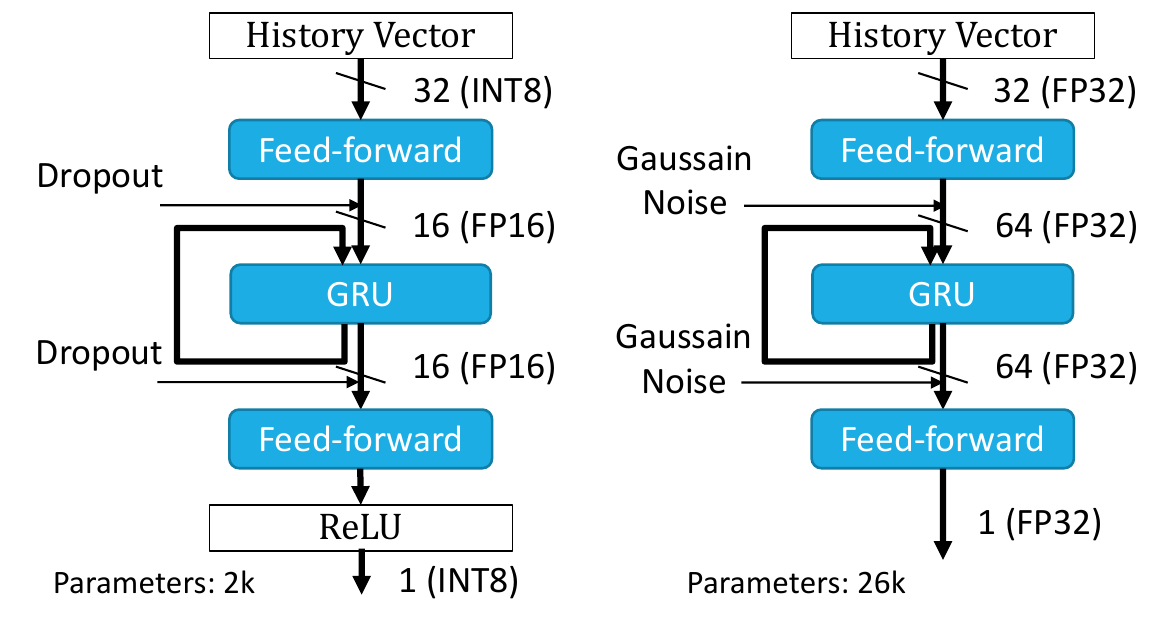}
\caption{Architectures of the defenders for the memory (left) and power (right) side channels.}
\label{fig_generator}
\end{figure}

The left side of Figure~\ref{fig_generator} shows the defender architecture to match that cost.
We choose a structure of 3-layer DNN  with a GRU~\cite{gru} layer in the middle.
While we only process the history of 32 samples at each time step, the memory of GRU enables handling longer sequential patterns.
As we can only add latencies to obfuscate the side-channel, ReLU is applied to the final output of the defender model.
While most of the activations and weights can be quantized to INT8, the memory cell of the GRU layer is quantized to FP16 to prevent accumulate precision loss through the feedback loop.
For a noise injection~\cite{gannoise} method, we choose Dropout~\cite{dropout} to further reduce the cost.
Using the existing estimation results~\cite{rethinkingfloat} scaled down~\cite{processscaling} to TSMC 7nm, the estimated power and area are 0.74W at maximum and 0.69mm$^2$.
The power consumption is proportional to the main memory utilization so that the average power should be far less than the estimated maximum.
Since direct training of a tiny DNN is hard, we train a bigger model with 160 hidden neurons with \modelname{} and compress it to the target architecture. 



\subsection{Defending Application Power Side Channel}

\begin{figure}[ht]
\centering
\includegraphics[width=0.9\linewidth]{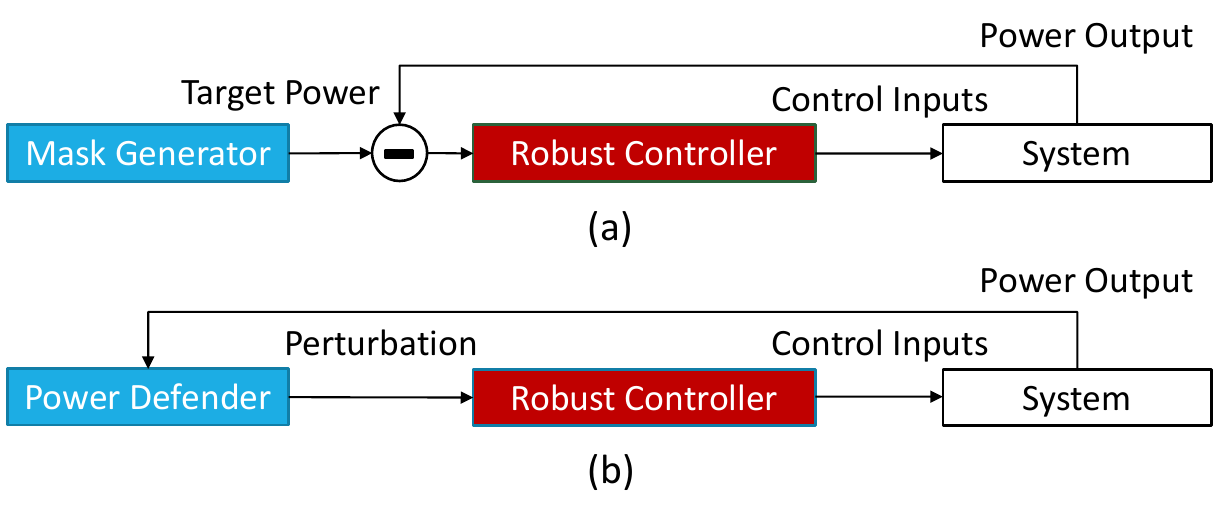}
\caption{Maya~\cite{maya} framework (a) and an ML defender plugged into the same framework (b).}
\label{fig_maya}
\end{figure}

The second case considers a chip multiprocessor running one of several victim applications.
The attacker periodically measures  the chip power consumption using the Intel RAPL interface~\cite{intelRapl}.
Based on the measurements over time, the attacker tries to deduce which application is running. 
We run PARSEC~\cite{parsec} applications so that the attack becomes a multi-label classification problem.

The side-channel operates at a slower speed because the fastest rate of RAPL measurements is once in a millisecond.
Hence, the envisioned defender is a software program.
The process measures the power consumed using RAPL, and spawns compute-intensive threads,
adds idle time, and changes the frequency
to distort the power consumed. The setup is based on the one described in Maya~\cite{maya}. 
Maya uses a randomized mask generator to distort the power trace; in this paper, we replace it with an ML defender, as shown in Figure~\ref{fig_maya}.

The defender's ML structure is illustrated at the right side of Figure~\ref{fig_generator}.
We can use a bigger DNN model with FP32 precision because we have $\approx 10^6 \times$ more time to process one inference compared to the memory side-channel defender.
The inference computation takes less than 1ms, which is negligible compared to Maya's 20ms sampling period.
An important difference in this deployment is that the defender must determine the noise for the \textit{next} time step, because it makes a decision only after measuring the current power.

\section{Experimental Methodology}
\label{setup}
\begin{figure}[t]
\centering
\includegraphics[width=\linewidth]{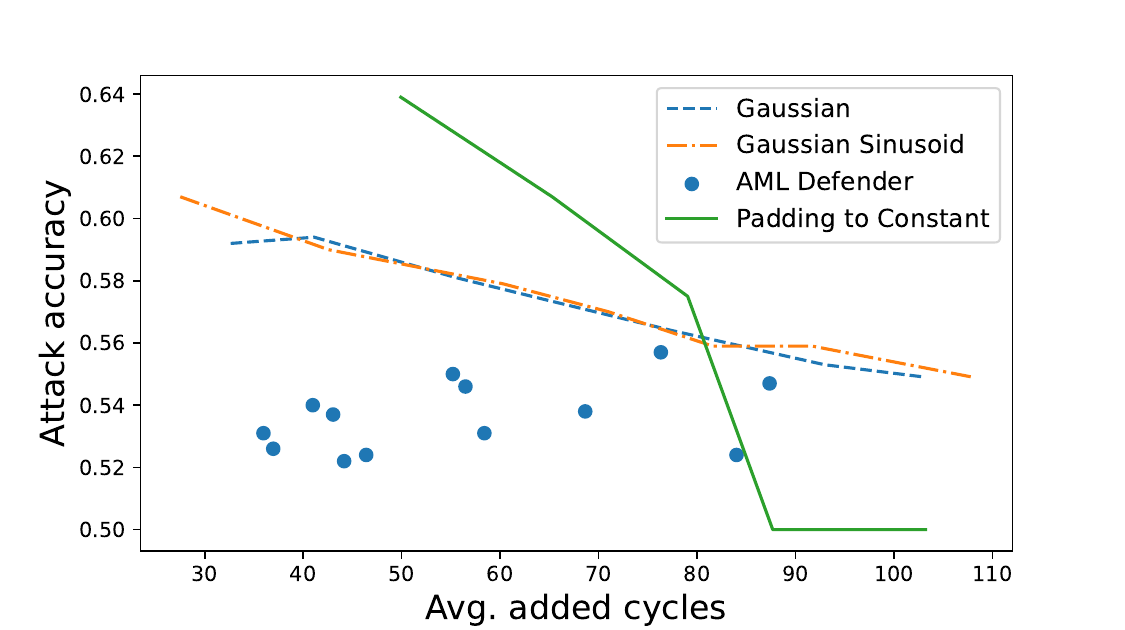}

\caption{Trade-off between security and slowdown of the noise generators for RSA. The x-axis is the average extra latency, while the y-axis is the accuracy of the best attacker.}
\vskip -0.1in
\label{fig_curve}

\end{figure}

\subsection{Data Collection}

\paragraph{Memory contention side-channel}
We base our attack code on the ring side-channel code of Paccagnella et al.~\cite{lordofthering}. We collect signal traces by running
two victim  crypto applications on a desktop with an Intel i5-9400 processor. 
One is the RSA decryption algorithm in \textit{libgcrypt}~\cite{libgcrypt} 1.5.2.
The other application is the EDDSA algorithm in the \textit{libgcrypt} 1.6.3.
The measurements are segmented into signals, each of which corresponds to a loop iteration of the target function. Each signal contains 42 and 105 samples for RSA and EDDSA, respectively.
To estimate the performance impact of our defender, we use the SST simulator~\cite{sst} modeling the desktop that generated the traces.
We run several memory-bound SPEC 2017 applications from~\cite{disintegratedcontrol}, to see the performance impact of higher memory latency.

\paragraph{Application power side-channel}
For this attack, we use the codebase of Maya~\cite{maya}.
We collect power traces on a desktop with an Intel Xeon W-2245 processor using RAPL~\cite{intelRapl}.
As victim applications, we use 10 PARSEC 3.0 \cite{parsec} applicatons (blackscholes,
bodytrack, canneal, freqmine, vips, streamcluster, splash2x.radiosity, splash2x.volrend, splash2x.water\_nsquared, and splash2x.water\_spatial) with simlarge inputs.
Each PARSEC signal is 500-samples long
to record the longest workloads (10 seconds).
The AML defender is trained using PyTorch and deployed to a real system using the Maya codebase and
the LibTorch API.

\subsection{Machine Learning Models}
\label{types_classifiers}

For the classifier in the \modelname{}, we use 1-D CNNs, which typically show the best attack accuracies.
For the  discriminator, we use an MLP, producing one 
real number for binary discrimination and $N$ numbers for $N$-class discrimination.
See Section~\ref{sec:sidechannel} for defender designs.

To perform our transferability analysis, we consider 7   ML classifiers.
Five of them are DNNs implemented using PyTorch:
a default 16-layer CNN  (CNN), a 25-layer CNN (CNN-Deep), a 16-layer CNN with double hidden neurons (CNN-Wide), a GRU-based
one (RNN), and an MLP (MLP).
The other 2 are non-DNNs implemented with the scikit-learn library: an SVM and a
K-Nearest Neighbor (KNN).

Additional implementation details (data collection, environmental setup, and evaluation methods) are available at the appendix and the source code in supplementary materials.

\section{Memory Side-channel Defense Evaluation}
\label{memoryeval}

\subsection{Security Provided}

\begin{table*}[t]
\caption{Attacker accuracy of seven classifiers as they attack three variations of our ML defender that were trained with 
the CNN classifier. No Discriminator is an uncompressed model trained without the discriminator in \modelname{}.}
    \label{table_transfer}
    \vskip 0.1in
\centering
\setlength{\tabcolsep}{8pt}
\renewcommand{\arraystretch}{1.1}
\begin{tabular}{|r|c|c|c|c|c|c|c|}
\hline
\multirow{2}{*}{Defender} & \multicolumn{7}{c|}{Classifier Accuracy}                      \\ \cline{2-8}
                           & MLP   & RNN   & CNN    & CNN-Deep & CNN-Wide & SVM   & KNN                             \\ \hline
Uncompressed                   & 0.513 & 0.507 & 0.507  & 0.506    & 0.507    & \textbf{0.524} & 0.513                    \\ \hline
No Discriminator           & 0.507 & 0.503 & 0.516  & 0.513    & 0.511    & \textbf{0.537}      & 0.528                                \\ \hline
Compressed                 & 0.526 & 0.520 & \textbf{0.532} & 0.528    & 0.528    & 0.531 & 0.511                    \\ \hline

\end{tabular}
\vskip -0.1in
\end{table*}

We evaluate how efficiently our ML defender secures the memory contention side-channel in terms of the security over noise level.
Here, the noise level is counted by the average extra delay in CPU cycles.
We consider three different obfuscation methods as baselines.
Gaussian and Gaussian Sinusoid are noise generator methods from Maya~\cite{maya} that generate random signal targets.
Since we cannot apply negative noise to the channel, the noise is applied only when the target is higher than the measured latency.
Another baseline is Padding to Constant which sets a minimum constant latency and delays all transactions to take at least that latency.
We adjust the average noise level by modifying their hyperparameters such as noise average and constant value.
The defender's noise level can also be adjusted by changing the L1 hinge value while training it.

Figure~\ref{fig_curve} shows the accuracy of the attack for each defender and level of noise---showing only the data
for the highest accuracy of the 7 classifiers. 
Note that the accuracy of 0.5 is the target for a perfect defense because the crypto key attack is a binary classification problem.
Padding to Constant is the only design that can achieve perfect defense---but only if we allow a large perturbation level of 88 average cycles or more.
The randomized generators (Gaussian and Gaussian Sinusoid) are not effective, especially at low average latencies.
Finally, our ML defender is the most secure for medium to low average latencies.
For the best defender (i.e., 43 average cycle latency), it can be shown that the randomized generators leak 12 times more bits 
per measurement, with the proportionally lower time to reconstruct the secret. 
This defender is used for further evaluations.

\begin{table*}[ht]
\caption{IPCs of applications with defensive ML or padding to constant defense  normalized to
the IPCs without defenses.} 
\label{table_perf}
\vskip 0.1in
\renewcommand{\arraystretch}{1.1}
\centering
\begin{tabular}{|r|c|c|c|c|c|c|c|c|c|}
\hline
                   & GCC   & Bwaves & MCF   & cactuBSSN & LBM   & Leela & Fotonik3D & XZ    & Geomean \\ \hline
ML Defender & 0.986 & 0.869  & 0.906 & 0.969     & 0.875 & 0.912 & 0.994     & 0.853 & \textbf{0.919}   \\ \hline
Padding to Constant    & 0.913 & 0.759  & 0.812 & 0.878     & 0.768 & 0.817 & 0.982     & 0.737 & \textbf{0.829}   \\ \hline
\end{tabular}
\end{table*}

\subsection{Transferability to Other Classifiers}
\label{sec:memory_transferability}

Our ML defender is trained with a CNN classifier and a discriminator. We now assess its transferability to 7 classifiers as shown in Table~\ref{table_transfer}.  
We consider three variations of our defender: the uncompressed design (160 hidden neurons),
the uncompressed design trained without a discriminator (No Discriminator), and the compressed and quantized design.
 
The range of accuracy numbers in the table is 0.51-0.54. This means that the generated perturbations 
are transferable to various types of attackers. We also see that 
CNNs with more layers (CNN-Deep) or hidden neurons (CNN-Wide) do not show better accuracy compared to the original CNN.
This observation suggests that using reasonably bigger DNN is unlikely to break the defense.
In addition, the existence of the discriminator during training typically helps the defender achieve better transferability, especially to non-DNN classifiers (SVM and KNN). This is despite the fact that, with discriminator, the average 
perturbation of the defender is smaller (43 versus 51 cycles).
Finally, compressing and quantizing the model exposes slightly more leakage, but the model is still transferable to a variety of classifier types.
Overall, these results suggest that our defender is likely to be transferable to other types of classifiers.

\subsection{Different Victim Application}

We now generate the defender for the EDDSA application as described in Section~\ref{amldefense}.
Hereby we maintain the same defender architecture and only the parameters of the defender are different.
Specifically, Table~\ref{table_victimchange} shows the 
attacker accuracies of the best classifiers against generators trained on one victim application
({\em Training}) but used to defend  another victim ({\em Target}).
This shows that we can reuse the same defender hardware for different victim applications by only re-configuring the parameters.

\begin{table}[ht]
\caption{Attacker accuracies against generators trained on one victim ({\em Training}) 
but used to defend  another victim ({\em Target}), along with their average added cycles.}\label{table_victimchange}
\vskip 0.1in
\renewcommand{\arraystretch}{1.1}
\centering
\begin{tabular}{|r|c|c|c|}
\hline
\multirow{2}{*}{Training Victim} & \multicolumn{2}{c|}{Target Victim} & \multirow{2}{*}{Avg. Cycles} \\ \cline{2-3}
                                 & RSA              & EDDSA           &                                                                       \\ \hline
RSA                              & 0.532            & 0.566           & 43                                                                    \\ \hline
EDDSA                            & 0.661            & 0.543          & 38                                                                    \\ \hline
\end{tabular}
\vskip -0.1in
\end{table}

If we train for RSA and defend EDDSA, we attain some level of protection. 
This means that our AML defender is at least partially transferable to other victim applications.
However, if we train for EDDSA and defend RSA, the attacker accuracy is high (0.661).
This data suggests that the perturbation patterns for short signals (e.g., RSA with 42 samples) may be applicable 
to longer signals (e.g., EDDSA with 105 samples), but not the other way.

\subsection{Performance Overhead}


We compare the performance overhead of activating our ML defender to that of applying the
Padding to Constant defense with an average number of cycles of 88---i.e., the cycles for which it
attains perfect protection. We run the memory-intensive SPEC applications with these defenses
and measure the average Instructions Per Cycle (IPC). Table~\ref{table_perf} shows the resulting
IPCs normalized to the IPCs without defenses. 
From the table, we see that the ML defender and 
Padding to Constant attain 8\% and 17\% relative slowdown, respectively, than with no defense.
This overhead is largely proportional to the average latency added per access (43 and 88,
respectively). Overall, Defensive ML limits the performance overhead of securing the side-channel.

\section{Power Side-channel Defense Evaluation}
\label{powereval}
We compare three defense environments under the Maya framework of Figure~\ref{fig_maya}: 
Gaussian Sinusoid (GaussSine) mask, our ML defender mask, and no defense. For each of them, 
Table~\ref{table_poweracc} shows the attack accuracy of different classifiers as they classify 10 PARSEC applications. Since there are 10 applications, an attack accuracy of 10\% corresponds to a perfect
defense. The table only shows a few representative classifiers, including CNN, which 
consistently shows the highest attack accuracy.

\begin{table}[ht]
\caption{Attack accuracy of different classifiers as they classify 10 PARSEC applications under three
different defense schemes. The table also shows the  application power.}\label{table_poweracc}
\vskip 0.1in
\centering
\begin{tabular}{|r|ccc|c|}
\hline
\multirow{2}{*}{Defense} & \multicolumn{3}{c|}{Classifier}                                             & \multirow{2}{*}{\begin{tabular}[c]{@{}c@{}}Average\\ Power\end{tabular}} \\ \cline{2-4}
                                                                          & \multicolumn{1}{c|}{SVM}    & \multicolumn{1}{c|}{MLP}    & CNN             &                                                                          \\ \hline
None                                                                      & \multicolumn{1}{l|}{72.9\%} & \multicolumn{1}{l|}{65.3\%} & \textbf{91.7\%} & 48W                                                                      \\ \hline
GaussSine                                                                        & \multicolumn{1}{l|}{11.3\%} & \multicolumn{1}{l|}{11.4\%} & \textbf{22.4\%} & 95W                                                                      \\ \hline
Defender (Ours)                                                                & \multicolumn{1}{l|}{10.5\%} & \multicolumn{1}{l|}{12.3\%} & \textbf{16.8\%} & 66W                                                                      \\ \hline
\end{tabular}
\vskip -0.1in
\end{table}

The table shows that the ML defender has the lowest attack accuracy. 
Theoretically, GaussSine should leak less than the ML defender because its perturbations are
completely independent of the victim application.
However, using our defensive ML approach provides better protection in practice.
We believe that the reason is that ML training causes the defender to create smaller
perturbations than GaussSine. As a result, controller deviations from
the mask are likely to be smaller and, therefore, less likely to leak information.
Gaussian Sinusoid defends against the SVM and MLP classifiers, but leaks information
with the stronger CNN.

\begin{figure}[ht]
\centering
\includegraphics[width=\linewidth]{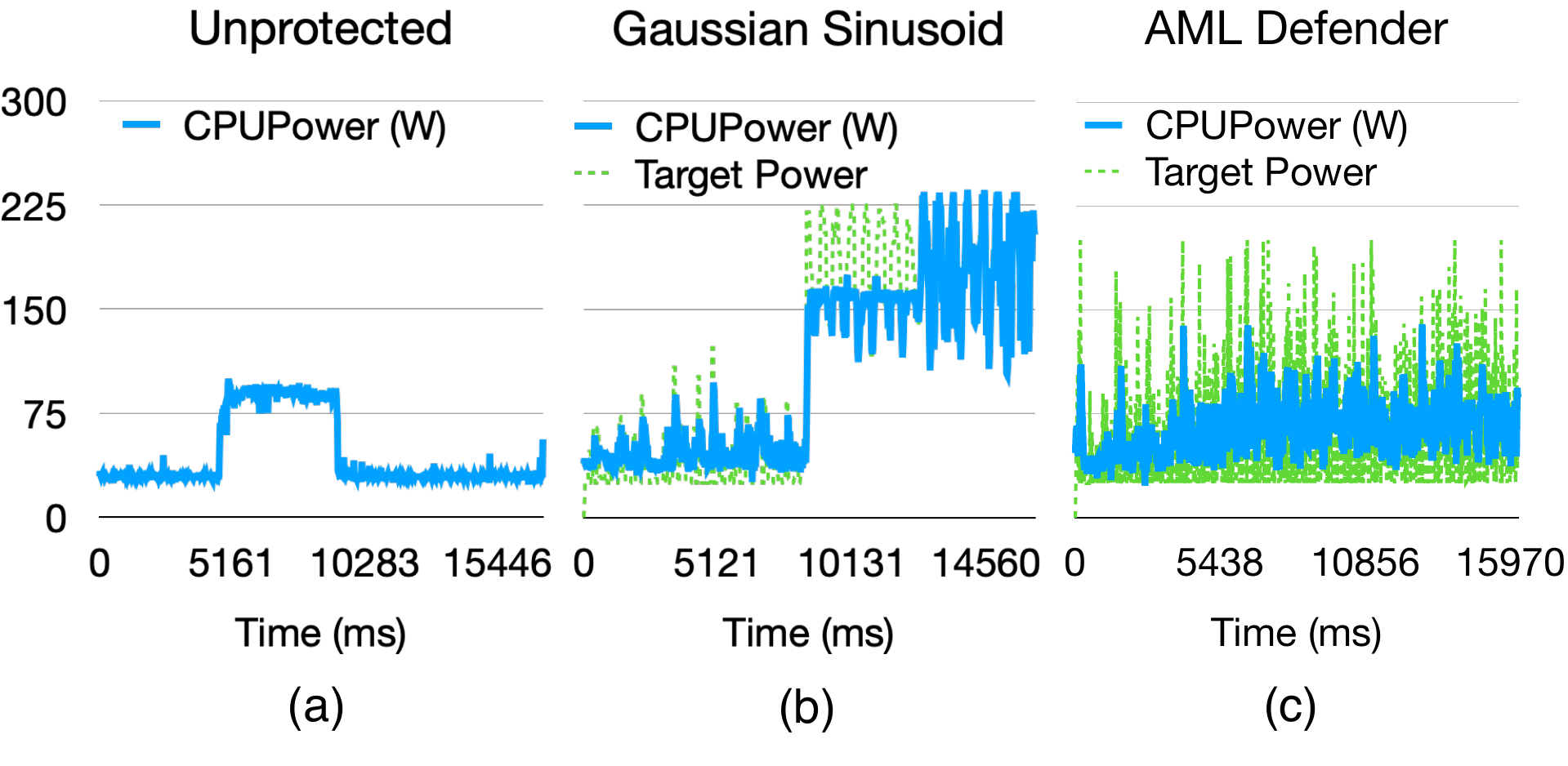}
\caption{Power traces running the PARSEC canneal benchmark. They are collected from the unprotected (a) and Maya-protected systems with different mask generators (b,c). Green dashed curves are power targets generated by the mask generators.
}
\label{fig:mayavsaml}
\vskip -0.1in

\end{figure}

The smaller perturbations in the defensive ML technique also result in lower application
power. The Gaussian Sinusoid mask generator creates power curves with wide power ranges.
On the other hand, the ML defender adjusts the power perturbations
depending on the application behavior; for applications with low power consumption,
the power perturbations are small. As a result, the average application power under
defensive ML is 66W, which is 70\% of the average power under Gaussian Sinusoid (95W),
and closer to the average power under no defense (48W).

Figure~\ref{fig:mayavsaml} gives the intuition about why the ML defender can be a better mask generator than the Gaussian Sinusoid.
The target power determined by the mask generator is enforced by Maya's robust controller by changing software-level knobs.
As seen in Figure~\ref{fig:mayavsaml}b, the controller may fail to follow the target if there is some application activity.
Maya's Gaussian Sinusoid generator creates long-range patterns that are independent of the application activity.
If the application activity falls into such a long period, the duration of the application can be leaked by observing the pattern.
This is not an issue for the ML defender, as the pattern dynamically changes depending on the observed power numbers.
It becomes much harder to pin the duration of the application when the ML defender is active on the system.

\section{Conclusion}
\label{conc_sec}

This paper showed that {\em Defensive} ML is a practical, general, 
and effective architectural technique
to  obfuscate signals from  
architectural side channels.
Our approach is based on strong ML methodology
and the ability to operate dynamically. 
We proposed a multi-network structure called \modelname{} to train  defenders adversarially.
We showed a workflow to design, implement, 
train, and deploy  transferable  defenders for different environments. We successfully applied
defensive ML to thwart two very different side-channel attacks: one
based on memory contention  and the other on application power. 
Future work will involve applying defensive ML to more environments.


\bibliography{references}
\bibliographystyle{icml2023}



\end{document}